\documentclass[aps,reprint,prl]{revtex4-1}
\usepackage{graphicx}
\usepackage{subfigure}
\usepackage{epstopdf, epsfig}

\usepackage{amsmath}
\usepackage{dcolumn}
\usepackage{bm}
\usepackage[colorlinks=true, linkcolor=blue]{hyperref}

\begin{document}

\title{On the role of magnetosonic solitons in perpendicular collisionless shock reformation}
\author{Renaud Gueroult}
\affiliation{Laplace, Universit\'{e} de Toulouse, CNRS, INPT, UPS, 31062 Toulouse, France}
\author{Yukiharu Ohsawa}
\affiliation{Department of Physics, Nagoya University, Nagoya 464-8602, Japan}
\author{Nathaniel J. Fisch}
\affiliation{Princeton Plasma Physics Laboratory, Princeton University, Princeton, NJ 08543, USA}

\date{\today}

\begin{abstract}
The nature of the magnetic structure arising from ion specular reflection in shock compression studies is examined by means of 1d particle in cell simulations. Propagation speed, field profiles and supporting currents for this magnetic structure are shown to be consistent with a magnetosonic soliton. Coincidentally, this structure and its evolution are typical of foot structures observed in perpendicular shock reformation. To reconcile these two observations, we propose, for the first time, that shock reformation can be explained as the result of the formation, growth and subsequent transition to a super-critical shock of a magnetosonic soliton. 
This argument is further supported by the remarkable agreement found between the period of the soliton evolution cycle and classical reformation results. This new result suggests that the unique properties of solitons can be used to shed new light on the long-standing issue of shock non-stationarity and its role on particle acceleration.

\end{abstract}


\maketitle


\paragraph*{Introduction.--}

Collisionless shocks have been intensively studied since the late 1950's by virtue of of the role they are believed to play in plasma heating and charged particle acceleration (see, \emph{e.~g.}, Refs.~\cite{Sagdeev1966,Biskamp1973,Blandford1987,Treumann2009,Burgess2012,Ohsawa2014,Marcowith2016} and references therein). One of the most important features in high Mach number shocks is the specular reflection of upstream ions, which serves as an energy dissipation mechanism~\cite{Kennel1985} to satisfy to shock conservation equations: ion specular reflection is paramount to both ion acceleration and shock structure. 


Further to its role in ion acceleration, ion specular reflection is responsible for the non-stationarity of quasi-perpendicular shocks. This temporal variability has been demonstrated through numerical simulations~(see, \emph{e.~g.}, Refs~\cite{Biskamp1972,Ohsawa1986,Lembege1987,Lembege2004,Shinohara2011}), observations~\cite{Lobzin2007,Mazelle2010,Sulaiman2015,Johlander2016} and experiments~\cite{Morse1972}. Although four different non-stationarity mechanisms have been suggested in full generality~\cite{Burgess2012}, the most likely candidate for explaining shock reformation in a 1-d exactly perpendicular shock ($\theta=90^{\circ}$,  with $\theta$ the angle between the upstream magnetic field and the shock normal) is the so-called \emph{self-reformation} mechanism (see, \emph{e. g.}, Ref.~\cite{Marcowith2016}). This self reformation cycle can be summarized as follows. First, ions reflected by the shock form a \emph{foot} ahead of the ramp. Due to the gyro-motion, these reflected ions pile up upstream of the foot at a distance slightly smaller than an ion gyro-radius ahead of the shock ramp, and create local magnetic field and density maxima there. Through a feedback mechanism, this foot then grows until it becomes as large as the initial shock ramp, effectively becoming the new ramp. Finally, this new ramp reflects incoming ions and the process repeats, with the shock advancing in a step-wise fashion. Numerical simulations suggest that the onset of this non-stationary reformation process is conditioned by large enough Mach number~\cite{Hada2003,Scholer2003,Burgess2007,Yuan2009} and fraction of incoming ions reflected by the shock~\cite{Hada2003}, and low enough ratio of plasma to magnetic pressure $\beta$~\cite{Hada2003,Scholer2003,Burgess2007,Yuan2009} and ion thermal velocity to shock velocity ratio~\cite{Scholer2003}. 

When non-stationarity conditions are met, the shock features change markedly over the course of a reformation cycle. Shock potential and shock ramp width display oscillations with a period of roughly $0.2-0.3$ times the upstream ion gyro-period~\cite{Nishimura2003,Kawashima2003,Treumann2009}. Consistent with these field oscillations, a complex and non-stationary ion dynamic is observed~\cite{Ohsawa1990,Chapman2005,Yang2009,Yang2009a,Yang2012}. 

Studying the interaction of an exploding plasma propagating through a background plasma, Yamauchi and Ohsawa~\cite{Yamauchi2007} showed that the magnetic deflection of an ion beam can lead to the formation of a magnetosonic pulse or \emph{soliton}. Identifying the strong similarities existing between this pulse and the magnetic bump formed over the course of the magnetic shock compression of a plasma channel~\cite{Gueroult2016}, Ohsawa conjectured~\cite{Ohsawa2016} that the magnetic bump observed in compression simulation results is in fact a magnetosonic pulse.

Solitons are one of the two kinds of stationary solutions to the Korteweg - de Vries (KdV) equation which can describe the propagation of weakly dispersive non-linear waves~\cite{Zabusky1965}. In particular, magnetosonic solitons are soliton solutions for the KdV equation derived for magnetosonic waves~\cite{Berezin1964,Kakutani1968}. A remarkable property of solitons is their stability with respect to interactions: they preserve their shape and speed after collision~\cite{Belashov2006}, behaving in some ways like particles. Beyond these unique physical properties, the concept of soliton enabled major theoretical developments in non-linear wave physics.


In this Letter, we report on the use of particle in cell (PIC) simulations to expose, for the first time, the role of magnetosonic solitons in the well-known shock self-reformation process in perpendicular shocks. We analyze the properties of the magnetic foot formed as a result of ion specular reflection at the shock front, confirming Ohsawa's conjecture~\cite{Ohsawa2016} and demonstrating that this foot actually evolves into a magnetosonic soliton. We also show that the growth of this soliton eventually leads to the formation of a super-critical shock, revealing the key role of this soliton in the shock reformation mechanism. Finally, we discuss the implications of these new findings.

\paragraph*{Numerical model.--}

\begin{figure}[h]
\begin{center}
\includegraphics[width=7.5cm]{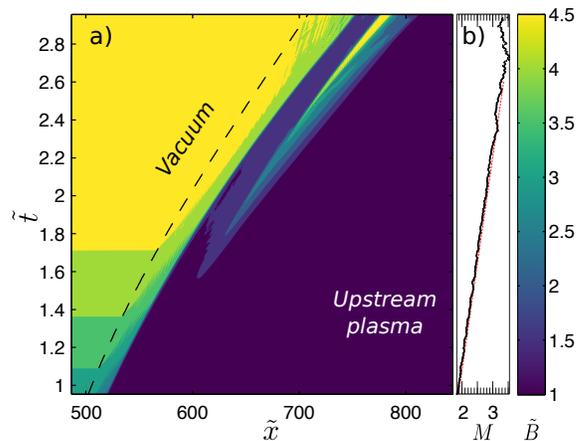}
\caption{a)~Normalized magnetic field contours. The upper left corner (left of the black dotted line) is the vacuum region. The lower right corner is the upstream plasma. The magnetic bump, visible as a narrow peak to the right of the shock front, grows with time. b)~Mach number \emph{vs}. time. $M$ increases almost linearly from $1.8$ to $3.5$ for $0.95\leq\tilde{t}\leq2.4$. Past $\tilde{t}\sim2.4$, $M$ begins to plateau. }
\label{Fig:BFieldContour}
\end{center}
\end{figure}

Simulations are carried out using the parallel fully electromagnetic relativistic PIC code Epoch~\cite{Arber2015}. A quarter sine magnetic compression ramp is generated at the left boundary of the 1d domain and propagates towards a pre-magnetized plasma slab, similarly to what was done in Ref.~\cite{Gueroult2016}. However, as opposed to this previous study, there is no symmetrical compression ramp generated at the right boundary in the simulation results presented in this Letter. When the compression ramp reaches the plasma slab, a shock wave develops and propagates to the right as illustrated in Figure~\ref{Fig:BFieldContour}. The computational domain is made of $2.5~10^{5}$ cells along $x$, with a spatial resolution $\Delta x$ of one cell per Debye length. The initial number of particles per cell per species is $20$. Both the upstream magnetic field $B_0$ and the bias magnetic field associated with the compression ramp are along $z$. Physical mass ratio $ m_e/m_p=\varepsilon^2 \sim 1/1836$ is used. The time step is $\Delta x/c$, with $c$ the speed of light. Dimensionless quantities are indicated by a tilde $\sim$, with time, length, speed and magnetic field normalized by the inverse of the upstream ion cyclotron frequency ${\Omega_{ci}}^{-1}$, the upstream electron skin depth $d_e = c/\omega_{pe}$, the upstream Alfven velocity $v_A$, and the upstream magnetic field $B_0$, respectively. The pre-magnetized plasma slab parameters, listed in Table~\ref{Table:Tab1}, correspond to the upstream plasma parameters.

\begin{table}
\begin{center}
\label{Table:Tab1}
\caption{Upstream dimensionless plasma parameters: thermal speed, Debye length, gyro-frequency, plasma frequency and Larmor radius. }
\begin{tabular}{c  c  c}
\hline
\hline
Parameter & Electrons & Protons\\
\hline
$\tilde{v}_{th}$ & $2.43$ & $0.057$\\
$\tilde{\lambda}_D$  & $0.0088$ & $0.0088$\\ 
$\tilde{\Omega}_{c}$ & $1.84~10^{3}$ & $1$\\
$\tilde{\omega}_{p}$ & $1.18~10^{4}$ & $2.75~10^{2}$\\
$\tilde{\rho}_{c}$ & $0.057$ & $2.43$\\
\hline
\hline
\end{tabular}
\end{center}
\end{table}

\begin{figure}[h]
\begin{center}
\includegraphics[width=7cm]{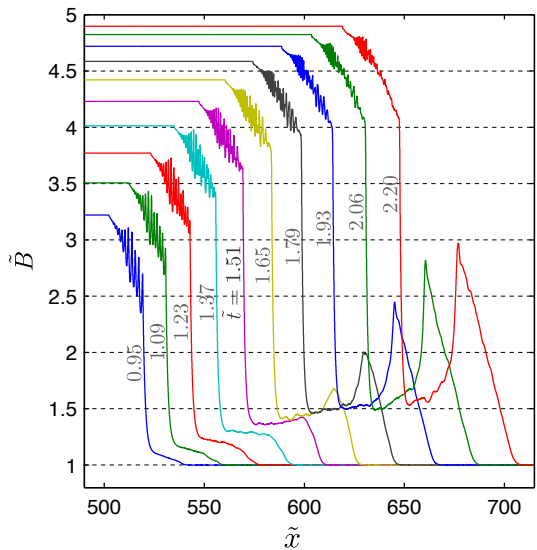}
\caption{Normalized magnetic field profiles at various times showing the onset and growth of a magnetic structure ahead of the shock front. }
\label{Fig:StackedBNoShift}
\end{center}
\end{figure}

\paragraph*{Foot formation and growth.--} Figure~\ref{Fig:StackedBNoShift} shows the onset and growth of the magnetic bump ahead of the shock ramp. Consistent with theory~\cite{Woods1969,Woods1971}, phase space distribution confirmed that the formation of this bump results from the specular reflection of upstream ions by the shock~\cite{Gueroult2016}.  For $\Delta \tilde{t}\sim0.7$ ($\tilde{t}\sim1.5$) after upstream ions began being specularly reflected by the shock, local magnetic field and density maxima are observed at the edge of the foot region. At this instant, the length of the foot is $\tilde{l}\sim 30$. Introducing $\rho_M$ the upstream gyro-radius of an ion specularly reflected with a velocity $Mv_A$, the foot length writes $\varepsilon \tilde{l} M^{-1}\rho_M$. Although the distance between the magnetic bump maximum and the shock ramp remains roughly equal to $30~d_e$ for $1.5\leq\tilde{t}\leq2.2$, the length of the foot grows to reach $\tilde{l}\sim60$ for $\tilde{t}\sim2.2$. For the $M\sim2$ shock found at the onset of specular reflection [see Figure~\ref{Fig:BFieldContour}a)], $\tilde{l}\sim60$ is equivalent to $0.7~\rho_M$, which is in remarkable agreement with the $0.68~\rho_M$ obtained from theory~\cite{Woods1971} and observation~\cite{Livesey1984} for a super-critical perpendicular shock. 

In view of the good agreement identified between simulation results and classical attributes of the magnetic foot which is commonly found in front of super-critical shocks, we further analyze the properties of this magnetic bump with the goal of shedding new light on the structure of super-critical perpendicular shocks.

\paragraph*{Magnetosonic soliton--} 

\begin{figure}[h]
\begin{center}
\includegraphics[width=7cm]{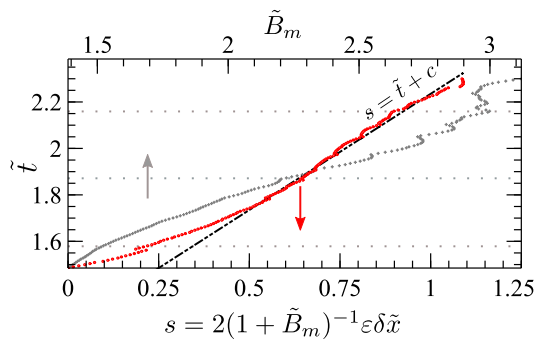}
\caption{Time evolution of the pulse's parameters. Red open circles show the re-normalized abscissa of the magnetic pulse maximum (lower horizontal axis), with $\delta \tilde{x} = \tilde{x}-\tilde{x}(\tilde{t}=1.485)$. Grey crosses indicate the pulse amplitude $\tilde{B}_m$ (upper horizontal axis). The black dash-dotted curve $s = \tilde{t}+c$ has a slope equal to $1$, typical of magnetosonic solitons.}
\label{Fig:MagnetosonicPulseSpeed}
\end{center}
\end{figure}

We first look at the phase velocity of the magnetic structure. Figure~\ref{Fig:MagnetosonicPulseSpeed} shows the time evolution of the position of the pulse maximum, along with the amplitude $\tilde{B}_m$ of the pulse. Here the maximum's abscissa is re-scaled, with $s=2(1+\tilde{B}_m)^{-1}\varepsilon\delta\tilde{x}$ and $\delta \tilde{x} = \tilde{x}-\tilde{x}(\tilde{t}=1.485)$. In the $(s,\tilde{t})$ coordinates used in Figure~\ref{Fig:MagnetosonicPulseSpeed}, we notice that the pulse displacement can be well fitted by a linear function of slope $1$ for $1.8\leq \tilde{t}\leq2.2$. In other words, the pulse's propagation speed is proportional to $(1+\tilde{B}_m)$ over this period, while $\tilde{B}_m$ grows from $1.8$ to roughly $3$. This scaling is characteristic of magnetosonic solitary pulses or solitons, which have a phase velocity $v_{\phi} = v_A(1+\tilde{B}_m)/2$~\cite{Adlam1958,Davis1958,Berezin1964,Gardner1965}. Note that this is also the propagation velocity for the transverse electric field, density and electric potential as the maxima of these different quantities in the pulse are collocated.

\begin{figure}[h]
\begin{center}
\includegraphics[width=7cm]{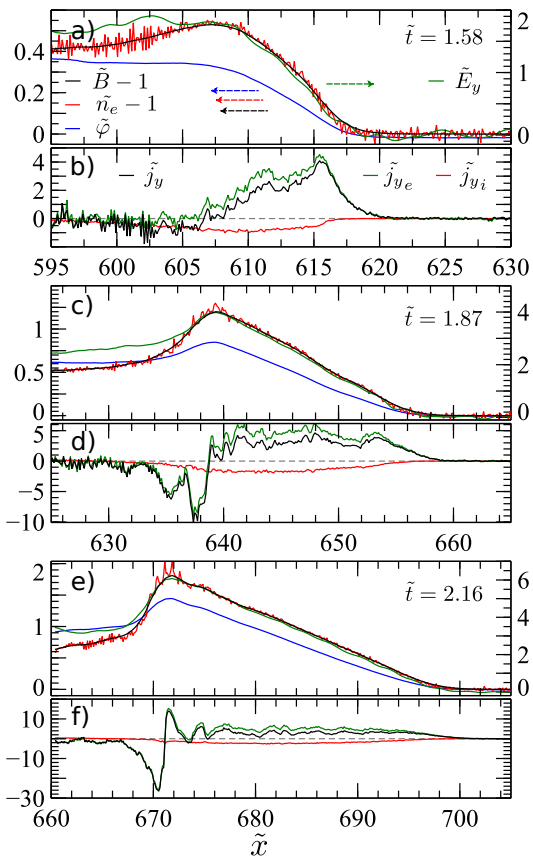}
\caption{Fields profiles at three different instants during the magnetosonic pulse growth: $\tilde{t} = 1.58$ a) and b), $\tilde{t} = 1.87$ c) and d) and $\tilde{t} = 2.16 $ e) and f). Normalized magnetic field (black), transverse electric field (green), electric potential (blue) and density (red) [a), c) and e)] and normalized electron (blue), ion (red) and total (black) transverse current densities [b), d) and f)]. }
\label{Fig:FieldEvolution}
\end{center}
\end{figure}

We now turn our attention to the profiles of normalized electric potential $e\tilde{\varphi} = e\varphi/(2m_p{v_A}^2)$, density $\tilde{n} = n/n_0$, transverse electric field $\tilde{E_y} = E_y/(v_A B_0)$ and transverse current densities $\tilde{j}_{y}$. First order expansion for the solutions of the Korteweg - de Vries (KdV) equation derived for perpendicular magnetosonic waves~\cite{Berezin1964,Kakutani1968} indicate analogous profiles for $\tilde{B}-1$, $\tilde{n}-1$, $\tilde{E}_y$ and $\tilde{\varphi}$. These quantities are plotted in Figures~\ref{Fig:FieldEvolution}a), \ref{Fig:FieldEvolution}c) and \ref{Fig:FieldEvolution}e) for three different instants during the pulse growth ($\tilde{t} = 1.58,1.87$ and $2.16$, shown by the horizontal grey dotted lines in Figure~\ref{Fig:MagnetosonicPulseSpeed}). Looking at Figure~\ref{Fig:FieldEvolution}, we indeed note a strong correlation between $\tilde{B}-1$, $\tilde{n}-1$, $\tilde{E}_y$ and $\tilde{\varphi}$, confirming the similarities with the field structure of a magnetosonic soliton. The density follows very closely the evolution of the magnetic field at all times, so that $B/n$ is nearly constant. The transverse electric field (plotted against the right vertical axis) scales roughly as $3.3(\tilde{B}-1)$ for $1.8\leq \tilde{t}\leq2.2$. Finally, the amplitude of the potential jump fits quite well to the theoretical value $\varphi_w = 2m_p{v_A}^2(M_w-1)/e$, with $M_w = (1+\tilde{B}_m)/2$ the Mach number of the pulse (here $M_w\sim 1.25, 1.6$ and $2$ for $\tilde{t} = 1.58,1.87$ and $2.16$), obtained for a non-linear magnetosonic pulse~\cite{Davis1958,Ohsawa1986a,Ohsawa2014}.


The evolution of transverse current densities, plotted in Figures ~\ref{Fig:FieldEvolution}b), \ref{Fig:FieldEvolution}d) and \ref{Fig:FieldEvolution}f) indicates that the current structure supporting the magnetic field bump is essentially an electron current, that is to say $\tilde{j}_{y}\sim\tilde{j}_{y_e}$. The small negative ion transverse current density $\tilde{j}_{y_i}$ is consistent with the magnetic deflection of specularly reflected ions. On the other hand, the electron current density $\tilde{j}_{y_e}$ exhibits peaks of opposite signs on either side of the magnetic field structure. This becomes particularly marked when the pulse is well formed ($\tilde{t} = 2.16$). This feature is typical of perpendicular magnetosonic solitons, of which the structure is determined by transverse electron currents under the hypothesis of charge-neutrality~\cite[p.156]{Ohsawa2014} 

From these three characteristics (propagation speed, self-similarities between $\tilde{B}-1$, $\tilde{n}-1$, $\tilde{E}_y$ and $\tilde{\varphi}$, and transverse currents), we conclude that the foot structure observed as a result of ion specular reflection by the shock ramp is a magnetosonic soliton.

\paragraph{Super-critical transition and reformation.--}

\begin{figure}[h]
\begin{center}
{\parbox{7cm}{\hspace{-0.cm}\includegraphics[]{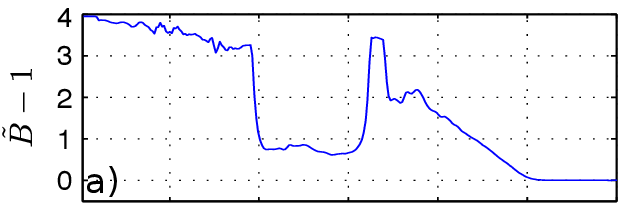}\\
\vspace{-0.1cm}\includegraphics[]{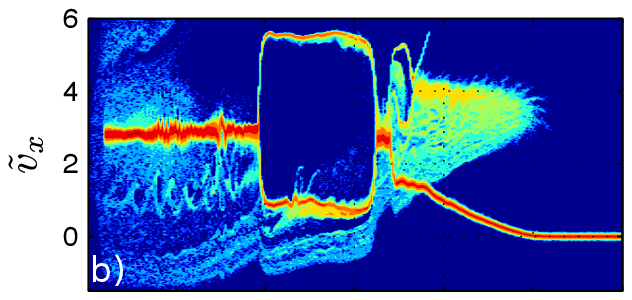}\\
\vspace{-0.15cm}\hspace{0.2cm}\includegraphics[]{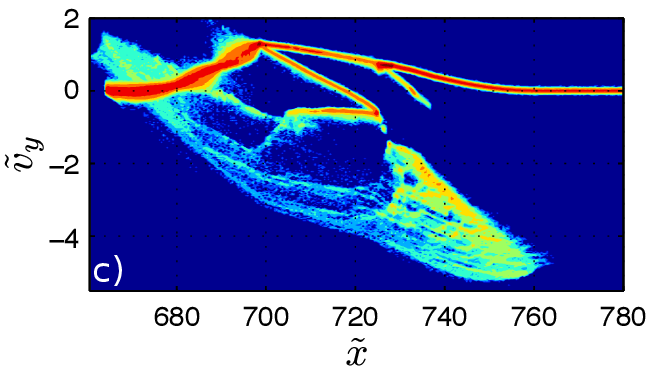}\\\vspace{0.15cm}}\label{Fig:Phase_Space4}
}
\caption{Magnetic field profile a) and ion phase space distribution [b) and c)] after the magnetosonic pulse became super-critical ($\tilde{t} = 2.59$). }
\label{Fig:PhaseSpace}
\end{center}
\end{figure}

When the pulse reaches $\tilde{B}_m\sim3$ (for $\tilde{t}\sim2.2$), a relatively sudden transition occurs, with the pulse amplitude growing rapidly to reach a value close to the one found downstream of the shock. This rapid growth takes place at the rear of the magnetic foot structure, as illustrated in Figure~\ref{Fig:PhaseSpace}a). Concurrent ion specular reflection by this new shock front is observed. This process is clearly visible in the form of a fast ion population ($\tilde{v}_x\sim5$ in the lab frame) in front of the pulse at $\tilde{t}=2.59$ in Figures~\ref{Fig:PhaseSpace}b) and \ref{Fig:PhaseSpace}c). The overgrown pulse then acts as the new shock front, quite similar to the step-wise advance commonly described in reforming shocks.


The fact that this transition and associated onset of specular reflection by the pulse are observed for $\tilde{B}_m\sim3$ provides further evidence for the formation of a magnetosonic soliton. As a matter of fact, the measured Mach number at the transition $M_w = (1+\tilde{B}_m)/2\sim2$ is exactly the critical Mach number obtained for a magnetosonic soliton with vanishing resistivity~\cite{Adlam1958,Davis1958,Auer1961,Biskamp1973}.

Finally, we notice (not shown here) that a new foot structure with a local maximum is formed for $\tilde{t}\sim3.44$ as a result of ion specular reflection by the pulse. By identification with the foot structure observed in front of the initial shock for $\tilde{t}\sim1.65$, and although only one cycle is simulated here, we infer a periodicity $\Delta\tilde{t}\sim1.8$, or $\Delta t\sim 0.3\times 2\pi/\Omega_{ci}$. Inasmuch as this value is in remarkable agreement with the period of a collisionless shock reformation cycle~\cite{Nishimura2003,Kawashima2003,Treumann2009}, we surmise that shock reformation is here mediated by the formation and subsequent growth beyond the critical Mach number of a magnetosonic soliton.

\paragraph{Conclusions.--}

We studied the formation of a magnetic pulse in fast compression experiments using 1d PIC simulations. We showed that this magnetic pulse, formed through ion specular reflection at the shock ramp, exhibits attributes common to the magnetic foot structure which is typically found in perpendicular collisionless shocks.

Through a detailed analysis of the properties of this pulse, in particular its propagation speed, its profile, and its supporting current structure, we inferred that it is actually a magnetosonic soliton. This soliton grows as it propagates upstream of the shock ramp and eventually reaches the critical Mach number $M\sim2$, whereupon it transitions to a super-critical shock. Thenceforth the overgrown pulse begins to reflect incoming ions, and serves as the new shock ramp. Finally, we showed that the period for this shock front step propagation through the formation and subsequent growth of a soliton is in remarkable agreement with the period of the well known self-reformation cycle in collisionless shocks.

In view of the strong similarities found between this mechanism and the self-reformation process, we surmise that our simulations reveal, for the first time, the mediating role of magnetosonic solitons in quasi-perpendicular shock reformation.

In showing that a large-amplitude pulse can create a new soliton due to kinetic and fluid effects, the present work brings to light novel facets of the evolution of magnetosonic pulses and of the creation of shock waves in astrophysical plasmas. In particular, these results pave the way for applying the large existing body of work on solitons to shed new light on the long-standing issue of shock non-stationarity and its role on particle acceleration. Incidentally, these results open new and promising perspectives for the use of fast magnetic compression in order to mitigate electron dephasing in plasma-based particle accelerators~\cite{Schmit2012}: plasma density profile near the beam axis could in principle be tailored from hollow to peaked-on-axis by taking advantage of the colliding properties of the counter-propagating solitons formed ahead of the shock.

\begin{acknowledgements}
This work was supported by US DOE Contract No. DE-SC0016072.
\end{acknowledgements}


%

\end{document}